\documentclass[english,aip,superscriptaddress,reprint]{revtex4-1}
\usepackage[T1]{fontenc}
\usepackage[latin9]{inputenc}
\usepackage{geometry}
\usepackage{amsmath}
\usepackage{amssymb}
\usepackage{graphicx}

\makeatletter
\usepackage{lipsum}

\makeatother

\usepackage{babel}
\begin{document}
\title{Lead Geometry and Transport Statistics in Molecular Junctions}
\author{Michael Ridley}
\affiliation{School of Chemistry, Tel Aviv University, Tel Aviv 69978, Israel}
\affiliation{The Raymond and Beverley Sackler Center for Computational Molecular
and Materials Science, Tel Aviv University, Tel Aviv 6997801, Israel}
\author{Emanuel Gull}
\affiliation{Department of Physics, University of Michigan, Ann Arbor, Michigan
48109, USA}
\affiliation{Center for Computational Quantum Physics, Flatiron Institute, New
York, New York 10010, USA}
\author{Guy Cohen}
\email{gcohen@tau.ac.il}

\affiliation{School of Chemistry, Tel Aviv University, Tel Aviv 69978, Israel}
\affiliation{The Raymond and Beverley Sackler Center for Computational Molecular
and Materials Science, Tel Aviv University, Tel Aviv 6997801, Israel}
\begin{abstract}
We present a numerically exact study of charge transport and its fluctuations
through a molecular junction driven out of equilibrium by a bias voltage,
using the Inchworm quantum Monte Carlo (iQMC) method. After showing
how the technique can be used to address any lead geometry, we concentrate
on one dimensional chains as an example. The finite bandwidth of the
leads is shown to affect transport properties in ways that cannot
be fully captured by quantum master equations: in particular, we reveal
an interaction-induced broadening of transport channels that is visible
at all voltages, and show how fluctuations of the current are a more
sensitive probe of this effect than the mean current.
\end{abstract}
\maketitle

\section{Introduction}

The conductance properties of single molecule junctions paint a rich
picture of the internal dynamics of open quantum systems far from
equilibrium. However, electron transport through a molecule coupled
to conducting leads at finite temperature is inherently stochastic,\citep{blanter_shot_2000}
and fluctuations in the current can provide a great deal more information
than the conductance or current alone.\citep{esposito_nonequilibrium_2009}
The current, its fluctuations or noise, and higher-order statistical
cumulants can be obtained from the so-called full counting statistics
(FCS) approach.\citep{levitov_charge_1993,levitov1996electron} In
recent years, noise and FCS techniques have been used to understand
many physical properties in both experimental and theoretical transport
settings. A few examples are quasiparticle charges in nanotube junctions;\citep{ferrier_universality_2016,ferrier_quantum_2017}
magnetic effects, channel determinations and thermotransport in atomic
junctions;\citep{vardimonExperimentalDeterminationConduction2013,kumar_shot_2013,vardimonOrbitalOriginElectrical2016,lumbroso_electronic_2018}
waiting time\citep{albert_distributions_2011,kosov2017waiting} and
first passage time\citep{ridley_numerically_2018} distributions;
correlations between currents in different leads;\citep{ridley2017partition}
thermodynamic efficiency fluctuations;\citep{esposito_efficiency_2015}
and fluctuation\textendash dissipation relations.\citep{nakamura_nonequilibrium_2010,kung_irreversibility_2012,pekola_towards_2015}

The treatment of noninteracting electron transport and FCS is at an
advanced stage, and both analytical and numerical methods are available.\citep{levitov_charge_1993,buttiker_four-terminal_1986,croy2009propagation,tang_full-counting_2014,ridley2015current}
However, the inclusion of many-body interactions in even the simplest
impurity models, where a small interacting impurity or molecule is
coupled to large noninteracting leads, remains theoretically challenging.
A variety of numerical approaches exists, each with different advantages
and disadvantages depending on the particular model in question and
the physical parameters. These might include, \emph{e.g.}, a lead\textendash molecule
coupling strength $\Gamma$, a temperature $k_{B}T$, an applied voltage
$V$ and a local Coulomb interaction strength $U$.

In addition to this set of discrete energy scales, molecular conductance
also depends strongly on the geometry of the leads and the nature
of the coupling between the lead and the molecule, a phenomenon which
is often studied theoretically in the context of noninteracting\citep{feng2008current,yang2014transient,zelovich_moleculelead_2015}
and weakly correlated\citep{darancet_quantitative_2012} systems.
Transport experiments show large variation of conductance with the
respect to the detailed structure of the lead\textendash molecule
interface, in particular the chemistry of the anchoring group at the
interface.\citep{zotti2010revealing,venkataraman_single-molecule_2006,quek_amine-gold_2007,hong2012single}
However, in simplified theoretical treatments, and especially when
dealing with many-body systems, it is often convenient to neglect
the internal electronic structure of the leads. One common assumption
is the wide band limit approximation.\citep{nitzan_chemical_2006,ridley2017partition}
Assuming a finite lead bandwidth $D$ is perhaps the simplest concession
to experimental reality that might be made, and is enough to show
that a strong effect on transport exists even in the non-interacting
case.\citep{oz_evaluation_2018} A more general approach for interacting
systems is therefore highly desirable.

In the regime of small lead\textendash molecule coupling and high
temperatures, the quantum master equation (QME) method offers a fast
and elegant means for investigating the basic conduction properties\citep{harbola2006quantum,esposito2010self}
and FCS\citep{bagrets_full_2003,kiesslich2006counting} of molecular
junctions. The basic Lindblad or Redfield QME approaches assume Markovian
dynamics, thus neglecting memory effects due to finite bandwidth and
lead structure. Most QME methods are also perturbative in the molecule\textendash bath
coupling, and thus restricted to low-order effects like sequential
tunneling. More advanced generalizations of the QME method include,
for example, non-Markovian memory effects\citep{flindt2008counting}
and inelastic cotunneling processes.\citep{kaasbjerg_full_2015} Novel
formulations more loosely connected to the QME even allow for the
approximate treatment of strong correlations, and may also include
the effects of external driving fields.\citep{dou_born-oppenheimer_2017,dou_universal_2018}
Equation of motion approaches, also loosely related, can be formulated
so as to correctly capture the noninteracting limit\citep{zubarev_double-time_1960,levy_steady_2013,levy_symmetry_2013,levy_absence_2019};
and other approaches are exact at the semiclassical limit.\citep{swenson_application_2011,swenson_semiclassical_2012}
Still, many of these approximations lack some of the advantages of
more naive QME approaches, which may result, \emph{e.g.}, in the violation
of the positivity of the density matrix\citep{breuer2002theory} or
a failure to satisfy various detailed balance and fluctuation\textendash dissipation
relations.\citep{esposito2006fluctuation,derrida2007non} It is also
typically difficult to verify the accuracy of these and other approximate
methods or to improve them systematically.

More recent developments have lead to the emergence of approaches
which map the correlated impurity model onto an auxiliary Lindblad
quantum master equation where the dynamics of the impurity and a small
number ($\lesssim10$) of effective lead levels are treated at the
full many-body level, with this extended system coupled to an effective
Markovian environment. The auxiliary QME can then be solved using
exact diagonalization (ED).\citep{arrigoni_nonequilibrium_2013,dorda_auxiliary_2014,dorda_auxiliary_2015-1}
It is possible to greatly improve this scheme by performing a perturbative
expansion in the interaction that is built around the exact solution
of the extended model.\citep{chen_auxiliary_2019} Nevertheless, the
many-body Hilbert space grows exponentially with the size of the extended
system, and only a limited level of detail regarding the structure
of the lead can therefore be included at a high level of accuracy.

There is a great deal of interest in \emph{numerically exact} methods
in this context. While this term has different definitions in the
literature, here we call a method numerically exact (in some parameter
regime) if arbitrarily precise results with reliable confidence intervals
can be obtained at a computational scaling that is polynomial in the
precision. At this level, the numerical renormalization group (NRG)
method has been very successful at studying the equilibrium properties
of interacting impurity models,\citep{bulla2008numerical} and is
ideally suited for exploring low energy properties. Extensions beyond
equilibrium exist and work well in many cases,\citep{anders_real-time_2005,anders_numerical_2008,anders_steady-state_2008,pletyukhov2012nonequilibrium}
but often struggle with large voltages, long timescales and high-energy
nonequilibrium properties. Similar considerations apply to time-dependent
density matrix renormalization group (tDMRG) approaches\citep{dias_da_silva_transport_2008,langer_real-time_2009,heidrich-meisner_real-time_2009,wolf_solving_2014}
and multiconfiguration time-dependent Hartree\textendash Fock.\citep{wang_multilayer_2003,wang_numerically_2009,wilner_bistability_2013,wilner_nonequilibrium_2014,wilner_phonon_2014,wilner_sub-ohmic_2015,wang_multilayer_2018}
Promising recent advances combine some of these ideas with auxiliary
master equation approaches, allowing for calculations with \textasciitilde 10\textendash 20
auxiliary lead sites.\citep{schwarz_lindblad-driven_2016,fugger_nonequilibrium_2018,schwarz_nonequilibrium_2018}

The hierarchical equation of motion (HEOM) method\citep{tanimura_quantum_1991,jin_exact_2008,li_hierarchical_2012}
offers an alternative numerically exact scheme that is efficient for
regimes in which the coupling-to-temperature ratio $\Gamma/k_{B}T$
is small.\citep{hartle_decoherence_2013,hartle_transport_2015} This
method is different from most of the wavefunction approaches above
in that it considers truly infinite leads, but relies on representing
the lead density of states in terms of a sum of a small number of
Lorentzian functions, making it difficult to study finite or structured
bands; nevertheless, recent progress has made significant headway
towards structured leads and low temperatures.\citep{erpenbeck_extending_2018,shi_efficient_2018}

Another set of methods are based on iterative schemes for summing
path integrals\citep{makri_time-dependent_1999,segal_numerically_2010-1,simine_path-integral_2013,weiss_iterative_2008,eckel_comparative_2010,kilgour_path-integral_2019}
that allow for long time propagation and detailed leads. These methods
rely on a memory cutoff that makes them difficult to converge in correlated
regimes or in the presence of narrow bands or detailed structure in
the leads, all of which lead to non-Markovian effects.\citep{werner_weak-coupling_2010,cohen_memory_2011,cohen_generalized_2013,cohen_numerically_2013}

In recent years, continuous time quantum Monte Carlo (CTQMC) techniques\citep{gull_continuous-time_2011}
have made great progress in addressing many of the limitations mentioned
above. A primary advantage is that, these being Green's functions
methods, it is straightforward to take into account an extremely large
number of bath sites or even to directly take the continuum limit.
Calculations involving $\sim10^{6}$ lead sites or directly at the
continuum limit are routinely performed, and the size or structure
of the leads is not a bottleneck in practice. CTQMC is usually formulated
in equilibrium and imaginary time; this is because real time implementations
able to address quantum transport\citep{muhlbacher_real-time_2008,schiro_real-time_2009,werner_diagrammatic_2009,schiro_real-time_2010,werner_weak-coupling_2010,cohen_memory_2011,cohen_numerically_2013,gull_numerically_2011,cohen_greens_2014-1,cohen_greens_2014,antipov_voltage_2016}
suffer from a dynamical sign problem. This refers to the exponential
growth in the error when stochastically sampling from a number of
Feynman diagrams that grows exponentially with increasing simulation
time. However, the dynamical sign problem can often be effectively
bypassed by an ``Inchworm'' algorithm (iQMC) that replaces a single
simulation of the full time propagation with a series of propagation
steps, each of which recycles information from previously calculated
propagation steps to shorter times.\citep{cohen_taming_2015,chen_inchworm_2017,chen_inchworm_2017-1,antipov_currents_2017,dong_quantum_2017,boag_inclusion-exclusion_2018}
In this regard, we note that several other interesting and promising
approaches to real time quantum Monte Carlo have recently been introduced.\citep{profumo_quantum_2015,polyakov_stochastic_2017,moutenet_cancellation_2019,bertrand_quantum_2019,bertrand_reconstructing_2019,kubiczek_exact_2019}
In practice, obtaining numerically exact data from iQMC relies on
converging the data in stochastic noise, a time discretization, and
usually also a maximum diagram order. The procedures for establishing
convergence, as well as the detailed scaling properties and error
analysis methodologies, have been discussed at length in the literature.\citep{cohen_taming_2015,antipov_currents_2017,boag_inclusion-exclusion_2018,cai_inchworm_2018}

Crucially for the present work, the computational efficiency of the
iQMC approach does not directly depend on the lead structure. iQMC
is therefore a promising approach to the study of quantum transport
through structured molecular junctions. Furthermore, the iQMC approach
was recently applied to the evaluation of FCS, allowing for preliminary
studies of shot noise in the Coulomb blockade and Kondo regimes.\citep{ridley_numerically_2018}
It is therefore also possible to address noise and higher moments
of the particle transport statistics in these regimes.

In this paper, we describe how iQMC can be used to study the effect
of lead geometry and strong many-body interactions on the steady state
current and noise characteristics of molecular junctions. We focus
on the effect of the band width in a regime where the QME might be
expected to perform reasonably well at the wide band limit. It is
shown that while QME is able to provide us with intuition even at
finite band width, its ability to provide accurate results rapidly
breaks down at lower band widths. The rest of the paper proceeds as
follows: Section~\ref{sec:Model} describes our model. In Section~\ref{subsec:Inchworm-qmc-method}
we briefly outline the iQMC approach to FCS and the calculation of
cumulants within this approach. We also briefly describe the QME approach
to FCS in Section~\ref{subsec:Master-equations-method}. In Section~\ref{subsec:Coupling-density-method}
we provide a practical and general numerical scheme for treating lead
structure. Two different types of leads with different dimensionality
are then considered in Section~\ref{subsec:Lead-geometry,-coupling}.
The general physics from the QME viewpoint is discussed in Section~\ref{subsec:General-picture-qme}.
The main results of the paper, showing both iQMC and QME data for
the current and noise dependence on voltage for a range of different
bandwidths, are presented in Section~\ref{subsec:Master-equations-vs.}.
Section~\ref{sec:Conclusions} contains our conclusions.

\section{Model\label{sec:Model}}

We consider a real-space realization of the nonequilibrium Anderson
impurity model, a simple phenomenological model of a magnetic impurity
that is often used to explore transport through a molecular or atomic
junction in the presence of local electron\textendash electron interactions.
The Hamiltonian is given by
\begin{equation}
\hat{H}=\hat{H}_{M}+\sum_{\ell}\hat{h}_{\ell}+\sum_{\ell}\hat{h}_{M\ell}.
\end{equation}
Here, the isolated molecular subspace $M$ is assumed to contain a
single spin-degenerate level. It is described by the interacting local
Hamiltonian
\begin{equation}
\hat{H}_{M}=\sum_{\sigma}\varepsilon^{\sigma}\hat{d}_{\sigma}^{\dagger}\hat{d}_{\sigma}+U\hat{d}_{\uparrow}^{\dagger}\hat{d}_{\uparrow}\hat{d}_{\downarrow}^{\dagger}\hat{d}_{\downarrow},
\end{equation}
where $\sigma\in\left\{ \uparrow,\downarrow\right\} $ is a spin index
and $\varepsilon^{\sigma}$ is the energy needed to introduce a single
electron of spin $\sigma$ onto the molecule. The $\hat{d}_{\sigma}^{\dagger}\left(\hat{d}_{\sigma}\right)$
are creation (annihilation) operators for a spin-$\sigma$ electron
on the molecule, and $U$ represents the interaction energy between
two electrons simultaneously occupying the junction. The molecule
is coupled to several infinite sets of noninteracting tight-binding
sites comprising the leads. In particular, the term
\begin{equation}
\hat{h}_{\ell}=\sum_{\sigma}\sum_{i\in\ell}\varepsilon_{i}^{\sigma}\hat{a}_{\sigma i}^{\dagger}\hat{a}_{\sigma i}+\sum_{\sigma}\sum_{i\neq j\in\ell}t_{ij}^{\sigma}\hat{a}_{\sigma i}^{\dagger}\hat{a}_{\sigma j}
\end{equation}
describes lead $\ell$ in terms of on-site energies $\varepsilon_{i}^{\sigma}$
and hopping energies $t_{ij}^{\sigma}$. The sites $i$ can be thought
of as localized atomic orbitals. The molecule\textendash lead coupling
is given by the linear coupling term
\begin{equation}
\hat{h}_{M\ell}=\sum_{\sigma}\sum_{i\in\ell}\left(t_{0i}^{\sigma}\hat{a}_{\sigma0}^{\dagger}\hat{a}_{\sigma i}+\mathrm{H.C.}\right).
\end{equation}

Our method is agnostic towards the specific details of the leads (\emph{i.e.}
the structure of $t_{ij}^{\sigma}$ and $\varepsilon_{i}^{\sigma}$),
but we note that we have assumed that these parameters are diagonal
in spin; removing this constraint would require a generalization of
the model.\citep{hartle_decoherence_2013} The dynamics of physical
observables in the model above when starting from a factorized initial
condition can then be evaluated in terms of its dependence on the
coupling density
\begin{equation}
\Gamma_{\ell\sigma}\left(\omega\right)=\pi\sum_{k\in\ell}\left|t_{0k}^{\sigma}\right|^{2}\delta\left(\omega-\varepsilon_{k}^{\sigma}\right),\label{eq:coupling_density_definition}
\end{equation}
where the single-particle lead Hamiltonian $\hat{h}_{\ell}$ is diagonal
in the $k$ basis, \emph{i.e.} $\hat{h}_{\ell}=\sum_{\sigma}\sum_{k}\varepsilon_{k}^{\sigma}\hat{a}_{k}^{\dagger}\hat{a}_{k}$.
Since $\hat{h}_{\ell}$ is noninteracting (in the sense that it is
quadratic), the diagonalization can sometimes be carried out analytically
and can easily be carried out numerically for leads comprising thousands
of sites. If the hopping elements $t_{ij}^{\sigma}$ are local in
space, it can easily be carried out numerically for millions of sites
using sparse matrix techniques, and we describe how to do this below.
This allows for treating a rich variety of physical systems.

\section{Methodology\label{sec:Methodology}}

\subsection{Inchworm quantum Monte Carlo\label{subsec:Inchworm-qmc-method}}

The iQMC method can be used to simulate the dynamics of a many-body
quantum system prepared in some initial state and propagated in time
by some Hamiltonian. The nonequilibrium steady state which forms when
the system in question is open is extracted from dynamical information
at the long time limit. In a coupling quench or partitioned\citep{ridley2015current}
approach, the system is initially in the decoupled state $\rho_{0}=\rho_{M}\otimes\prod_{\ell}\rho_{\ell}$,
a stationary state of $\hat{H}_{0}\equiv\hat{H}_{M}+\sum_{\ell}\hat{h}_{\ell}$.
At time zero, the coupling or hybridization term $\sum_{\ell}\hat{h}_{M\ell}$
is activated and particles can begin to flow between the molecule
and the leads. A nonequilibrium steady state will eventually develop
if the leads are chosen to be infinite and if the thermodynamic parameters
(\emph{e.g.} temperature or chemical potential) of different leads
are taken to be different from each other. We assume that the molecular
density matrix $\rho_{M}$ is prepared in one of the four eigenstates
$\left|\phi_{i}\right\rangle \left\langle \phi_{i}\right|$ of $\hat{H}_{M}$,
where $\hat{H}_{M}\left|\phi_{i}\right\rangle =E_{i}\left|\phi_{i}\right\rangle $.
In what follows we will focus on steady state quantities that do not
depend on the choice of initial local eigenstate. Whereas most work
done so far uses this partitioned method, where we propagate only
along the two real-time (Keldysh) branches of the contour, it is also
possible to include propagation along an imaginary time branch and
introduce the voltage at the initial time. This corresponds to a voltage
quench or partition-free version of the inchworm method.\citep{antipov_currents_2017}
We expect the partition-free approach to be more efficient at exploring
equilibrium and very small voltages, but it turns out to be more computationally
expensive in the present context.

In the original iQMC approach, one evaluates the propagator

\begin{equation}
p_{\phi\phi'}\left(z_{1},z_{2}\right)\equiv\left\langle \phi\right|\textrm{Tr}_{\left\{ \ell\right\} }\Bigg[\rho_{0}T_{\mathcal{C}}e^{-i\int_{z_{1}}^{z_{2}}\mathrm{d}z\,\hat{H}\left(z\right)}\Biggr]\left|\phi'\right\rangle \label{eq:iQMC_prop}
\end{equation}
between two time points $z_{1},\thinspace z_{2}$ on the Keldysh contour,
where $\phi$ and $\phi'$ denote different atomic states for the
initial preparation and $T_{\mathcal{C}}$ is the contour time-ordering
operator. This is done by expanding $p_{\phi\phi'}$ in powers of
the hybridization and summing the resulting diagrammatic series stochastically.
Modified propagators describing specific observables are also evaluated.\citep{cohen_greens_2014}
Instead of summing over all possible diagrams on the contour in what
constitutes a brute force or ``bare'' Monte Carlo approach,\citep{werner_diagrammatic_2009}
the iQMC algorithm incrementally evaluates propagators on longer and
longer time intervals along the contour, while optimally reusing propagator
data computed on shorter intervals. This avoids the dynamical sign
problem in a wide variety of cases, and the scaling of the computational
cost with simulation time is effectively reduced from exponential
to quadratic.\citep{cohen_taming_2015} In practice, we typically
limit an additional diagram order parameter approximately corresponding
to the maximum order of the self-energy in a self-consistent diagrammatic
expansion; the method becomes inefficient if the order needed to obtain
convergence is very large,\citep{boag_inclusion-exclusion_2018} which
we expect to happen when the underlying expansion does not capture
the physics of the parameter regime under investigation.\citep{chen_inchworm_2017-1}

Recently, the iQMC technique was extended to the study of FCS of particle
transport.\citep{ridley_numerically_2018} Rather than evaluating
the propagator in Eq.~\ref{eq:iQMC_prop}, we evaluate the moment
generating function for some lead $\ell$,

\begin{equation}
Z_{\ell}\left(\lambda,t\right)\equiv\underset{\Delta n_{\ell}}{\sum}P\left(\Delta n_{\ell},t\right)e^{i\lambda\Delta n_{\ell}}.\label{eq:Z_def}
\end{equation}
Here, $P\left(\Delta n_{\ell},t\right)$ is the probability that the
number of particles in lead $\ell$ has changed by $\Delta n_{\ell}$
at time $t$. The expression for $Z\left(\lambda,t\right)$ can be
reformulated as a propagator $p_{\phi\phi'}^{\lambda}$ in which the
Hamiltonian is modified by a counting field $\lambda$.\citep{esposito_nonequilibrium_2009,tang_full-counting_2014,ridley_numerically_2018}
The parameter $\lambda$ changes sign as the time variable crosses
the folding point on the Keldysh contour and the subsequent modification
to the Hamiltonian only affects the hybridization term, $\hat{h}_{M\ell}\left(z\right)\rightarrow\hat{h}_{M\ell}\left(\lambda;z\right)$.\citep{tang_full-counting_2014,ridley_numerically_2018}

The technical details regarding the iQMC algorithm and the evaluation
of $Z\left(\lambda,t\right)$ including all modifications to the iQMC
algorithm are discussed at length elsewhere.\citep{cohen_taming_2015,antipov_currents_2017,ridley_numerically_2018}
For the purposes of the present work we merely note that the $k^{\mathrm{th}}$
order statistical cumulant is extracted from $k^{\mathrm{th}}$ order
logarithmic derivative of $Z_{\ell}\left(\lambda,t\right)$ with respect
to $\lambda$:

\begin{equation}
C_{\ell,k}\left(t\right)=\frac{\partial^{k}\log\left(Z_{\ell}\textrm{\ensuremath{\left(\lambda,t\right)}}\right)}{\partial\left(i\lambda\right)^{k}}.\label{eq:cumulant_def}
\end{equation}
The steady state current and current noise in lead $\ell$ are then
evaluated from the asymptotic gradients of the first and second order
cumulants: 

\begin{align}
I_{\ell} & =\underset{t\rightarrow\infty}{\lim}\frac{C_{1,\ell}\left(t\right)}{t},\label{eq:current_SS}\\
S_{\ell} & =\underset{t\rightarrow\infty}{\lim}\frac{C_{2,\ell}\left(t\right)}{t}.\label{eq:noise_SS}
\end{align}
In iQMC, since the logarithmic derivatives are performed numerically
over noisy data, the extraction of specific cumulants becomes increasingly
difficult as their order grows.

\subsection{Master equations\label{subsec:Master-equations-method}}

In this section, we briefly outline the Markovian QME treatment that
we use to compare with the iQMC method and as a guide to our exploration
of the parameter space. The QME provides an approximate expression
for the dynamics of the reduced density matrix describing the embedded
molecule, $\hat{\sigma}\left(t\right)\equiv\mathrm{Tr}_{\left\{ \ell\right\} }\left\{ \hat{\rho}\left(t\right)\right\} $.
Here $\ell$ is the lead subspace, and $\hat{\sigma}$ is therefore
a $4\times4$ matrix in the many-body molecular basis $\left|\phi_{i}\right\rangle $.
Since the local Hamiltonian and the hybridization are diagonal in
this basis, the different site populations are not coupled through
the off-diagonal terms of $\hat{\sigma}\left(t\right)$ and we may
neglect coherences. We thus write down a simple equation of motion
for the four diagonal matrix elements $p_{i}=\left\langle \phi_{i}\right|\hat{\sigma}\left(t\right)\left|\phi_{i}\right\rangle $:
\begin{equation}
\frac{\mathrm{d}p_{i}\left(t\right)}{\mathrm{d}t}=\sum_{j}M_{ij}p_{j}\left(t\right).\label{eq:EoM}
\end{equation}
The matrix elements $M_{ij}\equiv\left\langle \phi_{i}\right|\hat{M}\left|\phi_{j}\right\rangle =\underset{\ell}{\sum}M_{ij}^{\ell}$
correspond to transition rates from dot state $j$ to state $i$ as
mediated by the lead $\ell$:

\begin{equation}
M_{ij}^{\ell}=\left\{ \begin{array}{cc}
-\underset{k}{\sum}R_{ki}^{\ell} & i=j,\\
R_{ij}^{\ell} & i\neq j.
\end{array}\right.\label{eq:M_rates}
\end{equation}
Here, the rates $R_{ij}^{\ell}$ are given by
\begin{equation}
R_{ij}^{\ell}=\left|\epsilon_{ij}\right|\Gamma_{\ell}\left(\epsilon_{ij}\Delta E_{ij}\right)f_{\ell}\left(\Delta E_{ij}\right).\label{eq:R_rates}
\end{equation}
These rates therefore depend on the coupling density $\Gamma_{\ell}\left(\omega\right)$
and the Fermi function $f_{\ell}\left(\omega\right)=\left(e^{\beta\left(\omega-\mu_{\ell}\right)}+1\right)^{-1}$,
as well as on the energy differences between molecular states $\Delta E_{ij}\equiv E_{i}-E_{j}$.
We also use the definition and $\epsilon_{ij}=\pm1$ for $n_{i}-n_{j}=\pm1$
and $\epsilon_{ij}=0$ otherwise, where $n_{i}$ is the number of
electrons on the molecule in the local state $\left|\phi_{i}\right\rangle $;
and the fact that $f_{\ell}\left(-\Delta E\right)=1-f_{\ell}\left(\Delta E\right)$.
Intuitively, the rate $R_{ij}^{\ell}$ for $\epsilon_{ij}=1$ describes
a process where the molecule goes from state $j$ to state $i$ by
absorbing an electron of energy $\Delta E_{ij}$ from an occupied
level in the leads, and the corresponding rate for $\epsilon_{ij}=-1$
describes a process where an electron of energy $\Delta E_{ji}=-\Delta E_{ij}$
is released into the leads.

In the presence of a nonzero counting field $\lambda$, the modified
populations $p_{i}\left(t\right)\rightarrow p_{i}\left(\lambda,t\right)$
are obtained by modifying the matrix elements in Eq.~\ref{eq:M_rates}
according to
\begin{align}
M_{ij}^{\ell} & \rightarrow M_{ij}^{\ell}e^{i\lambda\epsilon_{ij}}.\label{eq:M_lambda}
\end{align}
The generating function $Z\left(\lambda,t\right)$ is then obtained
as the trace with respect to the molecular subsystem,\citep{bagrets_full_2003,flindt2004full,esposito_nonequilibrium_2009}

\begin{equation}
Z\left(\lambda,t\right)=\mathrm{Tr}_{S}\left[\hat{\sigma}\left(\lambda,t\right)\right]=\sum_{i}p_{i}\left(\lambda,t\right),\label{eq:GF_QME}
\end{equation}
and all cumulants can be extracted as described in Section~\ref{subsec:Inchworm-qmc-method}.

\subsection{Coupling density\label{subsec:Coupling-density-method}}

For completeness, we will briefly review how the coupling density
Eq.~\ref{eq:coupling_density_definition} for a particular lead is
obtained in practice, given a concrete model of the system and leads.
To simplify the discussion and the notation we will drop the $\ell$
and $\sigma$ subscripts. We further assume (as is the case in the
rest of this work) that the molecule is coupled only to a single site
in the lead, such that 
\begin{equation}
t_{0i}=\delta_{i1}t_{M}.\label{eq:coupling_assumption}
\end{equation}
It is straightforward to generalize this to more complex models than
that used here; for example, analogous procedures for time-dependent
noninteracting transport and in the presence of secondary Markovian
leads have been discussed in detail in the literature.\citep{zelovich_state_2014,zelovich_moleculelead_2015,hod_driven_2016,zelovich_driven_2016,zelovich_parameter-free_2017}

We begin by rewriting Eq.~\ref{eq:coupling_density_definition} in
terms of Green's functions:
\begin{equation}
\Gamma\left(\omega\right)=\pi\sum_{k}\left|t_{0k}\right|^{2}\delta\left(\omega-\varepsilon_{k}\right)=\frac{1}{2}\sum_{k}\left|t_{0k}\right|^{2}A_{k}\left(\omega\right),\label{eq:Gamma_from_A}
\end{equation}
where $A_{k}\left(\omega\right)$ is the spectral function of the
noninteracting lead in the diagonal $k$ basis. This is given by the
imaginary part of the retarded Green's function $G_{k}^{R}\left(\omega\right)$:
\begin{equation}
\begin{aligned}A_{k}\left(\omega\right) & =-2\Im\left\{ G_{k}^{R}\left(\omega\right)\right\} ,\\
G_{ij}^{R}\left(\omega\right) & =\mathcal{F}\left\{ G_{ij}^{R}\left(t\right)\right\} ,\\
G_{ij}^{R}\left(t\right) & =i\vartheta\left(t\right)\left\langle \left[\hat{a}_{i}^{\dagger}\left(0\right),\hat{a}_{j}\left(t\right)\right]\right\rangle .
\end{aligned}
\end{equation}
Here, $\mathcal{F}$ is a Fourier transform from time to frequency
and $\vartheta\left(t\right)$ is the Heaviside step function. The
diagonalization of the lead subspace $\hat{h}$ (corresponding to
a single spin component of $\hat{h}_{\ell}$ above) is performed in
the single particle picture, \emph{i.e.}
\begin{equation}
\varepsilon_{k}=\sum_{ij}V_{ki}\left\langle i\right|\hat{h}\left|j\right\rangle V_{jk}^{\dagger},
\end{equation}
with $V_{ki}$ a unitary matrix whose dimension is the number of sites
in the lead, which we will denote as $N$. The states $\left|i\right\rangle =\hat{a}_{i}^{\dagger}\left|0\right\rangle $
are obtained by adding a single electron to the vacuum state $\left|0\right\rangle $.
The matrix $V_{ki}=\left.\left\langle k\right|i\right\rangle $ defines
the diagonal $k$ basis and its relationship to the real space $i,j$
basis. In particular, it allows us to make the transformations
\begin{equation}
\begin{aligned}A_{k}\left(\omega\right) & =\sum_{ij}V_{ki}A_{ij}\left(\omega\right)V_{jk}^{\dagger},\\
\left|t_{0k}\right|^{2} & =t_{k0}t_{0k}=\sum_{ij}V_{ki}t_{i0}t_{0j}V_{jk}^{\dagger}=\left|t_{M}^{2}\right|V_{k1}V_{1k}^{\dagger}.
\end{aligned}
\label{eq:k_transforms}
\end{equation}
In the last equality, we have used Eq.~\ref{eq:coupling_assumption}.
Plugging Eq.~\ref{eq:k_transforms} into Eq.~\ref{eq:Gamma_from_A}
we obtain
\begin{equation}
\begin{aligned}\Gamma\left(\omega\right) & =\frac{\left|t_{M}^{2}\right|}{2}\sum_{k}V_{k1}A_{k}\left(\omega\right)V_{1k}^{\dagger}\\
 & =\pi t_{M}^{2}A_{11}\left(\omega\right).
\end{aligned}
\label{eq:Gamma_from_real_space}
\end{equation}
Our task is therefore reduced to calculating the local spectral function
at a single site on the lead.

While this task might most intuitively be carried out simply by diagonalizing
$\left\langle i\right|\hat{h}\left|j\right\rangle $, one is often
interested in large leads. Unless the diagonalization can be carried
out analytically, numerical diagonalization of the $N\times N$ matrix
is needed. The computational time for this scales as $O\left(N^{3}\right)$
and the memory as $O\left(N^{2}\right)$. However, given short-ranged
hopping terms, the matrix $\left\langle i\right|\hat{h}\left|j\right\rangle $
is very sparse and therefore the diagonalization can benefit from
sparse matrix techniques, which reduce both the time and memory costs
to $O\left(N\right)$. Such techniques are well known in the quantum
transport literature and implemented in widely used packages such
as Kwant.\citep{groth_kwant:_2014} A particularly simple and useful
approach is based on the so-called kernel polynomial method,\citep{weise_kernel_2006}
and we briefly outline it here.

We rewrite Eq.~\ref{eq:Gamma_from_real_space} as
\begin{equation}
\begin{aligned}\Gamma\left(\omega\right) & =\frac{\left|t_{M}^{2}\right|}{2}\sum_{k}\left.\left\langle k\right|1\right\rangle A_{k}\left(\omega\right)\left.\left\langle 1\right|k\right\rangle \\
 & =\pi\left|t_{M}^{2}\right|\sum_{k}\left|\left.\left\langle 1\right|k\right\rangle \right|^{2}\delta\left(\omega-\varepsilon_{k}\right).
\end{aligned}
\label{eq:Gamma_for_KPM}
\end{equation}
Let us assume that we are working in a set of units such that $\Gamma\left(\omega\right)$
is nonzero only within a range of frequencies $-1<\omega<1$. For
any physical lead with a finite bandwidth, this constraint can be
enforced by finding the highest and lowest eigenvalue of $\hat{h}$,
for example with the Lanczos method, and then appropriately rescaling
$\hat{h}$. Given this, we can expand $\Gamma\left(\omega\right)$
in a series of Chebyshev polynomials $T_{n}\left(\omega\right)$:
\begin{equation}
\begin{aligned}\Gamma\left(\omega\right) & =\frac{1}{\pi\sqrt{1-x^{2}}}\left[\mu_{0}+2\sum_{n=1}^{\infty}\mu_{n}T_{n}\left(\omega\right)\right],\\
\mu_{n} & =\int_{-1}^{1}\Gamma\left(\omega\right)T_{n}\left(\omega\right)\mathrm{d}\omega.
\end{aligned}
\label{eq:chebyshev_expansion}
\end{equation}
Combining Eq.~\ref{eq:Gamma_for_KPM} and Eq.~\ref{eq:chebyshev_expansion}
gives
\begin{align}
\mu_{n} & =\pi\left|t_{M}^{2}\right|\sum_{k}\left|\left.\left\langle 1\right|k\right\rangle \right|^{2}T_{n}\left(\varepsilon_{k}\right)\nonumber \\
 & =\pi\left|t_{M}^{2}\right|\sum_{k}\left\langle 1\right|T_{n}\left(\hat{h}\right)\left|k\right\rangle \left.\left\langle k\right|1\right\rangle \\
 & =\pi\left|t_{M}^{2}\right|\left\langle 1\right|T_{n}\left(\hat{h}\right)\left|1\right\rangle .\nonumber 
\end{align}
The $n^{\mathrm{th}}$ moment is therefore given in terms of the action
of the $n^{\mathrm{th}}$ Chebyshev polynomial of the Hamiltonian
$\hat{h}$ on the state $\left|1\right\rangle $:
\begin{equation}
\begin{aligned}\mu_{n} & =\pi\left|t_{M}^{2}\right|\left.\left\langle 1\right|\alpha_{n}\right\rangle ,\\
\left|\alpha_{n}\right\rangle  & =T_{n}\left(\hat{h}\right)\left|1\right\rangle .
\end{aligned}
\end{equation}

Now, using the recurrence relations connecting the Chebyshev polynomials,
it is easy to show that
\begin{equation}
\begin{aligned}\left|\alpha_{0}\right\rangle  & =\left|1\right\rangle ,\\
\left|\alpha_{1}\right\rangle  & =\hat{h}\left|1\right\rangle ,\\
\left|\alpha_{n+1}\right\rangle  & =2\hat{h}\left|\alpha_{n}\right\rangle -\left|\alpha_{n-1}\right\rangle .
\end{aligned}
\label{eq:recurrence_relation}
\end{equation}
To obtain coefficients up to any finite $n$ it is therefore necessary
only to multiply a sequence of vectors in the single particle $\left|i\right\rangle $
basis by the sparse matrix $\hat{h}$, a numerical task which can
be accomplished in $O\left(N\right)$ steps, and apply Eq.~\ref{eq:recurrence_relation}
recursively. The expansion is stable, inexpensive to perform and converges
rapidly. In practice, it is usually numerically beneficial to convolve
the results by a kernel; for this particular case, a Lorentz kernel
with a small width parameter $\lambda$ should be used to maintain
causality. Standard techniques allow this to be done be modifying
the expansion coefficients such that $\mu_{n}\rightarrow g_{n}\mu_{n}$,
where $g_{n}=\frac{\sinh\left(\lambda\left(1-\frac{n}{N}\right)\right)}{\sinh\left(\lambda\right)}$.\citep{weise_kernel_2006}

\section{Results}

\subsection{Lead geometry, coupling density and bandwidth\label{subsec:Lead-geometry,-coupling}}

\begin{figure}
\includegraphics{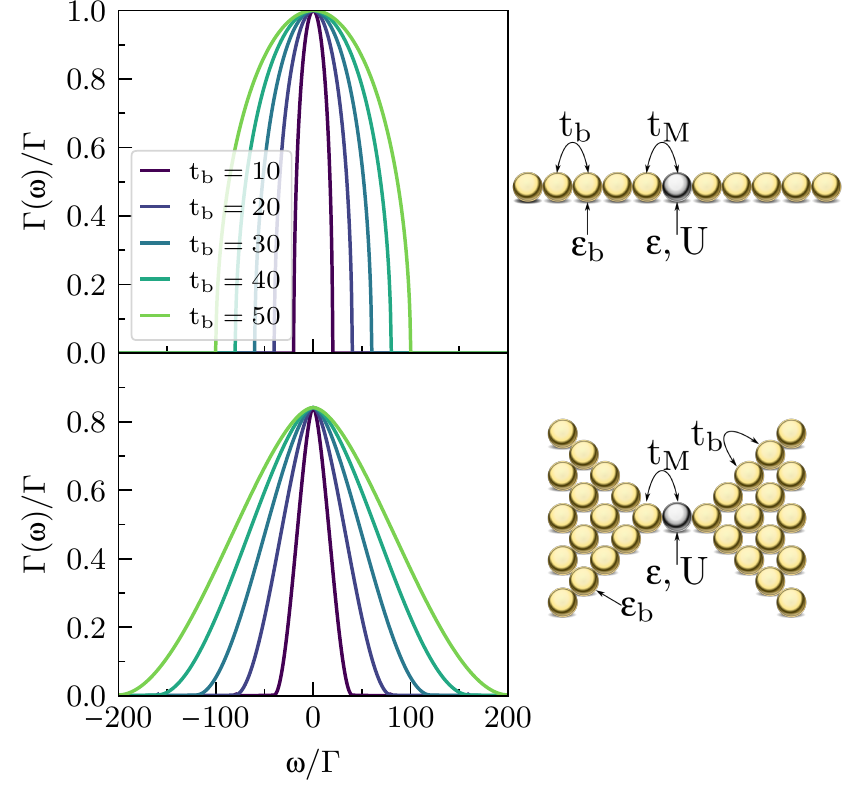}\caption{The coupling density $\Gamma_{\ell\sigma}\left(\omega\right)=\Gamma\left(\omega\right)$
for a single lead $\ell\in\left\{ L,R\right\} $ and spin $\sigma$
is is shown in the left panels for several values of $t_{b}$, given
two different types of leads illustrated in the right panels. On top,
a one-dimensional chain is considered (with scaling $t_{M}=\sqrt{t_{b}\Gamma}$).
On the bottom, a two-dimensional square lattice coupled to the molecule
at its corner is shown (with scaling $t_{M}=t_{b}$).\label{fig:model_and_coupling_density}}
\end{figure}

It is common and convenient in the study of molecular electronics
and other quantum transport problems to assume the wide band limit,
\begin{equation}
\Gamma_{\ell\sigma}^{\mathrm{wideband}}\left(\omega\right)\simeq\Gamma_{\ell\sigma}=\mathrm{const.}
\end{equation}
We will begin by briefly considering how and when this limit emerges
in nanoscale electronics. In Fig.~\ref{fig:model_and_coupling_density},
we consider two simple cases of molecular (or possibly atomic) junctions
with two leads, $\ell\in\left\{ L,R\right\} $. In the top part of
the figure, each lead is a one-dimensional chain of identical $s$-orbital
sites with nearest-neighbor couplings, such that $\varepsilon_{i}^{\sigma}=\varepsilon_{b\ell}$,
$t_{ij}^{\sigma}=t_{b}\delta_{i,j\pm1}$ and $t_{0i}^{\sigma}=\delta_{i1_{\ell}}t_{M}$,
where $1_{\ell}$ is the index of the site in lead $\ell$ adjacent
to the molecule (this is the Anderson\textendash Newns model\citep{newns1969dm}).
At the limit of an infinitely long chain, $\Gamma\left(\omega\right)$
can be evaluated analytically and forms an ellipse:
\begin{equation}
\Gamma_{\ell\sigma}^{1D}\left(\omega\right)=\begin{cases}
\frac{t_{M}^{2}}{2t_{b}^{2}}\sqrt{4t_{b}^{2}-\left(\omega-\varepsilon_{b\ell}\right)^{2}} & \left|\omega-\varepsilon_{b\ell}\right|\le2t_{b},\\
0 & \mathrm{otherwise.}
\end{cases}
\end{equation}
The value of $t_{b}$ sets the bandwidth $D=4t_{b}$, and the function
$\Gamma\left(\omega\right)$ attains its maximum value of $\frac{t_{M}^{2}}{t_{b}}$
when $\omega=\varepsilon_{b\ell}$. To simplify our exploration of
the parameter space, we set $\frac{t_{M}^{2}}{t_{b}}\equiv\Gamma$
to hold this maximum value constant and use $\Gamma$ as our standard
unit of energy; we therefore have $t_{M}=\sqrt{t_{b}\Gamma}$. The
result is plotted for a series of values of $t_{b}$ with $\varepsilon_{b\ell}\equiv\varepsilon_{b}=0$
in the top left panel of Fig.~\ref{fig:model_and_coupling_density}.
The same parameters will be used throughout most of the following
sections. To minimize the notation, we also set $\hbar=e=1$, such
that all units are given in terms of $\Gamma$. In addition, we choose
to perform calculations in the left lead, so that simplified notations
for current ($I\equiv I_{L}$) and noise ($S\equiv S_{L}$) may be
chosen.

It is clear that the wide band limit emerges when $t_{b}$ is significantly
larger than all other energy scales. It is not clear that this limit
should be generally applicable within molecular electronics. Importantly,
in the presence of strong local interactions $U$ or large potential
differences between the leads, the curvature at small frequencies
in Fig.~\ref{fig:model_and_coupling_density} suggests that significant
deviations from the wide band limit can be expected to occur even
when $t_{b}$ is $50$ times the magnitude of the maximal dot\textendash bath
coupling. This result does not strongly depend on our choice of parameter
scaling. The bandwidth, on the other hand, should be a dominant factor
only for smaller values of $t_{b}$.

In more concrete terms, our parameters range from the case where hopping
energies (or nearest-neighbor overlaps between orbitals at adjacent
atoms) within the metallic leads, $t_{b}$, are $\sqrt{10}\sim3.16$
times larger than the molecule\textendash lead hopping $t_{M}$; to
the case where $t_{b}$ is $\sqrt{50}\sim7.07$ times larger than
$t_{M}$. Such parameters could conceivably be realized in either
molecular electronics junctions or atomic junctions under strain.
For much weaker molecule\textendash lead coupling, the wide band limit
becomes appropriate and master equations may be expected to perform
well at high temperatures. When $t_{b}\sim t_{M}$, it is immediately
clear from the considerations above that master equations should not
be applied at any temperature, and we will not address this regime
here.

The one-dimensional case is extreme, and it would be prudent to take
into account the role of lead dimensionality. In the bottom part of
Fig.~\ref{fig:model_and_coupling_density} we therefore consider
a semi-infinite two-dimensional square lattice with its corner coupled
to the molecule. Here, $\varepsilon_{i}^{\sigma}=\varepsilon_{b\ell}$,
$t_{ij}^{\sigma}=t_{b}\delta_{i,j\in\mathrm{n.n.}}$ and $t_{0i}^{\sigma}=t_{M}\delta_{i1_{\ell}}$
where the site $i=1_{\ell}$ is at the corner of the lattice, adjacent
to the molecule. While the particular coupling density in this case
may perhaps be obtainable analytically (as are the one-dimensional
case above and, incidentally, the infinite-dimensional hypercube where
a Gaussian is obtained\citep{vollhardt_proceedings_1994}), we solve
for it numerically using the more general kernel polynomial scheme
outlined in section~\ref{subsec:Coupling-density-method}. One immediate
difference with respect to the one-dimensional case is in the scaling
needed to maintain a constant maximum value of $\Gamma\left(\omega\right)$:
here we take $t_{M}=t_{b}$. A second difference is in the bandwidth,
which is $D=8t_{b}$ instead of $4t_{b}$. Finally, the cutoff at
the band edge occurs much more smoothly in this case.

We reiterate that our iQMC method is compatible with any $\Gamma_{\ell\sigma}\left(\omega\right)$,
corresponding to any particular lead geometry, as an initial input.
Here, we will proceed to examine the one-dimensional Anderson\textendash Newns
chain within both the QME and iQMC methods, focusing on the effect
of the finite bandwidth. Interacting calculations for the 2D square
corner geometry and other higher-dimensional cases will not be presented
here.

\subsection{General picture within master equations\label{subsec:General-picture-qme}}

\begin{figure}
\includegraphics{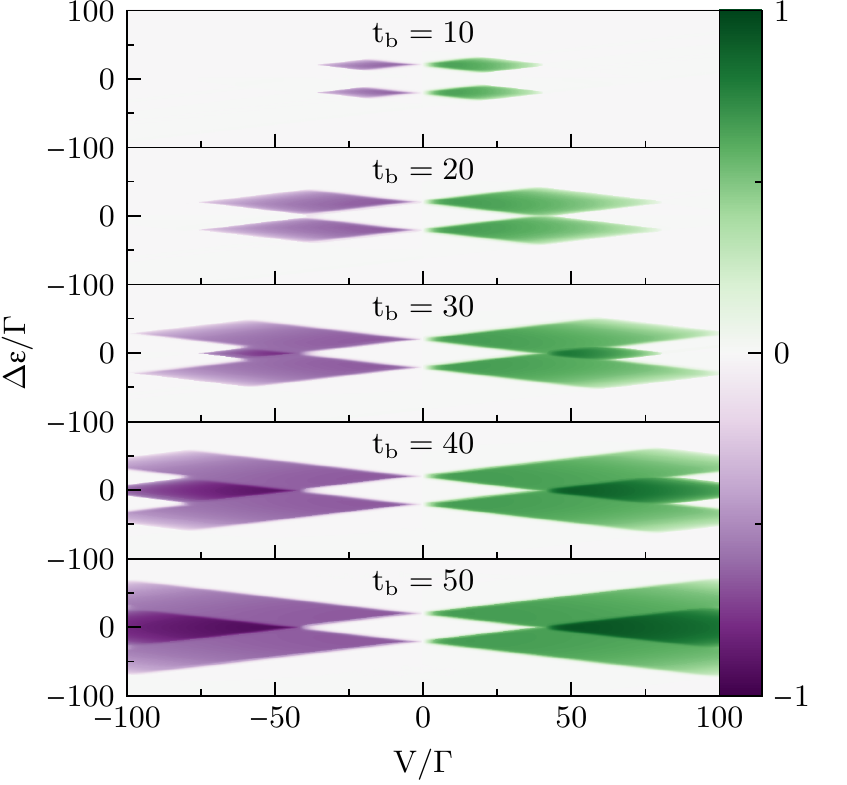}\caption{The master equation approximation to the steady state current $I$
in the chain is plotted as a function of the bias voltage $V=\mu_{L}-\mu_{R}$
and the gate voltage $\Delta\varepsilon$. The different panels show
$I\left(V,\Delta\varepsilon\right)$ at several values of the lead
hopping energy $t_{b}$, which determines the bandwidth $D=4t_{b}$.
The interaction strength is $U=40\Gamma$, the molecular level is
at $\varepsilon_{M}=-\frac{U}{2}+\Delta\varepsilon$ and the temperature
is $k_{B}T=\Gamma$.\label{fig:qme_contours}}
\end{figure}

Having established the general behavior of the coupling density within
the model, we continue to explore its effect on transport properties.
It is useful to start from an overall view connecting well-known Coulomb
blockade physics at the wide band limit (corresponding to a large
value of the hopping within the leads, $t_{b}$) with the behavior
at small band width (corresponding to small $t_{b}$). Furthermore,
we will consider parameters where $\Gamma$ is 1\textendash 2 orders
of magnitude smaller than every energy scale in the Hamiltonian, and
the temperature $k_{B}T$ is much larger than the Kondo temperature,
at which strongly correlated physics becomes important. At such parameters,
it is widely assumed in the literature that master equations can provide
us with at least a qualitatively correct picture. Specifically, we
choose interaction strength $U=40\Gamma$ and temperature $k_{B}T=\Gamma$,
far from the correlated regime; and apply bias voltages $V$ such
that the chemical potentials in the leads are $\mu_{\ell}=\varepsilon_{b\ell}=\frac{V}{2}$
for $\ell=L$ and $\mu_{\ell}=\varepsilon_{b\ell}=-\frac{V}{2}$ for
$\ell=R$. We also set the energy of half-filled states to $\varepsilon=-\frac{U}{2}+\Delta\varepsilon$,
such that the particle\textendash hole symmetric point is found at
$\Delta\varepsilon=0$. Experimentally, $\Delta\varepsilon$ corresponds
to a shift in the molecular potential or gate voltage caused by electrostatic
coupling to a third electrode.\citep{perrin2015single} In a typical
experiment a sweep of both the bias and gate voltages is performed
so that so-called stability diagrams can be constructed showing contour
plots of transport quantities as a function of both parameters.\citep{klein2004coulomb}

In Fig.~\ref{fig:qme_contours} we plot the steady state current
$I$ within the master equation approximation as a function of the
bias and gate voltages, $V$ and $\Delta\varepsilon$, at the same
series of $t_{b}$ values considered in Fig.~\ref{fig:model_and_coupling_density}.
At small $t_{b}$ (upper panel), the current is strongly suppressed
throughout the figure, whereas at large $t_{b}$ (lower panel) the
familiar Coulomb blockade diamond structure from the wide band limit\citep{perrin2015single}
is gradually restored.

The wide band picture, most closely approximated by the lowest panel
of Fig.~\ref{fig:qme_contours}, is characterized by\textemdash at
increasing voltage\textemdash regions with no current; regions with
where some of the conduction channels are open, and the current plateaus
at $I=\frac{2}{3}\Gamma$; and regions where all channels are open,
and the current plateaus at $I=\Gamma$.\citep{datta_electrical_2004}
As the bandwidth decreases (in increasingly higher panels) Finite
bandwidth resulting from the one-dimensional leads suppresses high
voltage transport in all channels, since no lead levels are available
to transfer electrons at higher energies. The small bandwidth more
strongly suppresses the opening of the second set of channels, which
are associated with higher energy transport processes. Therefore,
the dark triangles completely disappear at $t_{b}\lesssim20\Gamma$
and are smaller at $t_{b}\sim30\Gamma$. In particular, there exists
a region within the $t_{b}=30\Gamma$ plot (central panel) where at
voltages $V\simeq50\Gamma$ full and partial transport are separated
in $\Delta\varepsilon$ by a zero transport region. At slightly higher
voltages, full transport is completely suppressed whereas partial
transport (at certain values of $\Delta\varepsilon$) remains possible.

We note that it is possible to repeat this calculation with either
QME or iQMC for the two-dimensional leads (or any other leads), and
obtain results which differ quantitatively, since both the shape and
the width of the band will differ. However, as we are chiefly interested
in the qualitative effect of the limited bandwidth in the present
scope, we will continue to consider only the one-dimensional case.

It is now natural to ask whether the theoretical analysis above is
accurate. Temperature plays a central role here. Within the QME approximation,
the main effect of temperature is to smear out the borders between
the different conduction regions. At higher temperatures these borders
become smoother, and at lower temperatures they become sharper. It
is well understood that at very low temperatures the QME breaks down,
and that transport within the Kondo and mixed valence regimes is dominated
by entirely different mechanisms than those addressable by the QME.
However, at the parameters of Fig.~\ref{fig:qme_contours} the system
is almost three orders of magnitude above the Kondo temperature,\citep{hewson_kondo_1993}
and strong correlation effects should be largely irrelevant. Nevertheless,
the interaction is large enough that one might question the accuracy
of the single-particle picture in the leads after the activation of
the coupling. It is therefore of some interest to examine an exact
numerical solution of the model.

We emphasize that iQMC, the numerically exact method we will use,
is not restricted to high temperatures or weakly correlated physics.
In fact, in much of the work so far, iQMC has been used to study the
strongly correlated Kondo regime.\citep{cohen_taming_2015,antipov_currents_2017,dong_quantum_2017,ridley_numerically_2018,chen_auxiliary_2019}
We limit ourselves to high temperatures specifically so that any breakdown
of QME is unrelated to Kondo and therefore due only to the finite
nature of the band.

\subsection{Master equations vs. numerically exact iQMC results\label{subsec:Master-equations-vs.}}

\begin{figure}
\includegraphics{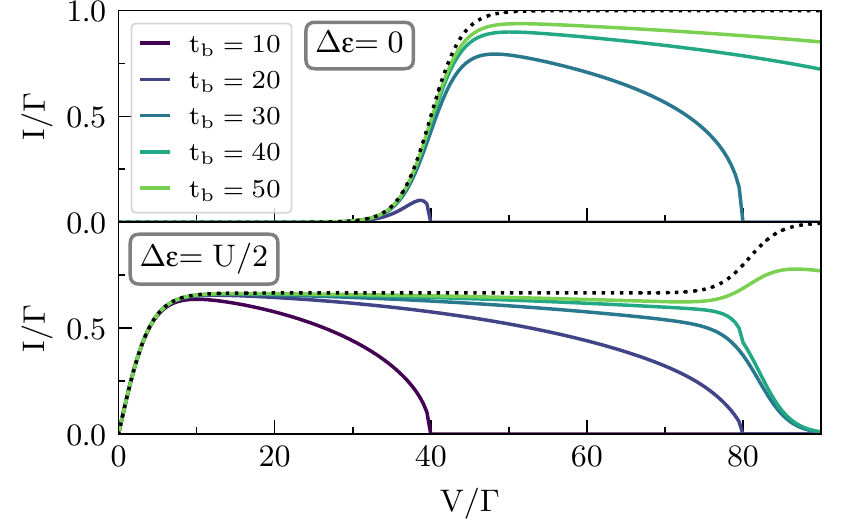}

\caption{The master equation approximation to the steady state current $I$
in the chain is shown as a function of bias voltage $V$, for gate
voltages at the symmetric point $\Delta\varepsilon=0$ (top) and at
$\Delta\varepsilon=\frac{U}{2}$ (bottom). The different curves in
each panel show several values of the lead hopping energy $t_{b}$,
which determines the bandwidth $D=4t_{b}$. The interaction strength
is $U=40\Gamma$, the molecular level is at $\varepsilon_{M}=-\frac{U}{2}+\Delta\varepsilon$
and the temperature is chosen such that  $k_{B}T=\Gamma$. As a guide
to the eye, the dotted curves in the two panels are the corresponding
master equation result at the wide band limit.\label{fig:qme_current}}
\end{figure}

\begin{figure}
\includegraphics{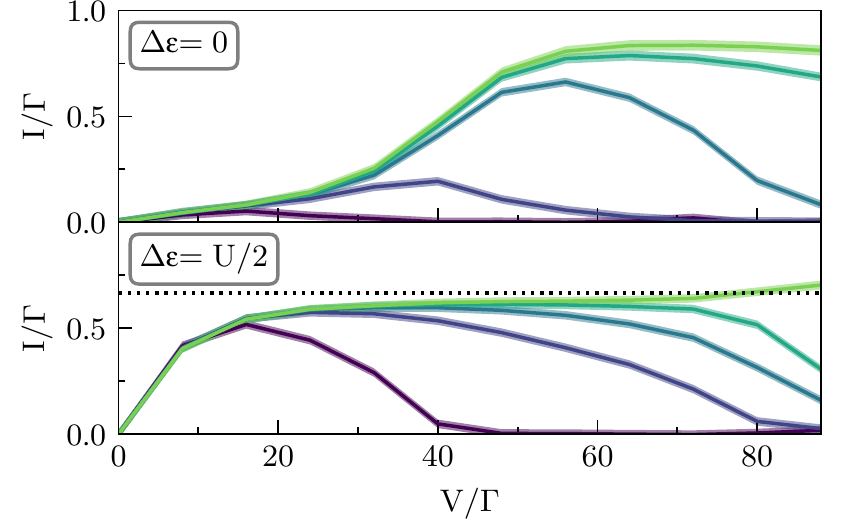}\caption{The numerically exact steady state current $I$ in the chain obtained
from iQMC is shown as a function of bias voltage $V$, for gate voltages
at the symmetric point $\Delta\varepsilon=0$ (top) and at $\Delta\varepsilon=\frac{U}{2}$
(bottom). Parameters are as in Fig.~\ref{fig:qme_current}.\label{fig:mc_current}}
\end{figure}

We now present our main results: an exploration of the phenomena predicted
by the master equation approximation in section~\ref{subsec:General-picture-qme}
within iQMC. In addition to the current $I$, we will also present
data for the shot noise $S$. To this end, we simulate the dynamics
of the generating function $Z\left(\lambda,t\right)$ up to a finite
time $t$ and for a constant value of $\lambda$, as described in
Sec.~\ref{subsec:Inchworm-qmc-method}. For the results shown here,
since we are far from the strongly correlated regime, to obtain reasonably
reliable results it is sufficient to go to time $t=\frac{2}{\Gamma}$
and to limit the maximum order of Inchworm diagrams to 3, corresponding
to a 2-crossing approximation. For some of the more difficult parameters,
we verified convergence by considering both longer times and higher
maximum orders (not shown). This is in contrast with the strongly
correlated regime of this model, where it is often necessary to propagate
to substantially longer times and orders.\citep{nuss_steady-state_2013,cohen_taming_2015,boag_inclusion-exclusion_2018,chen_auxiliary_2019}
Kondo physics can also result in very slow spin dynamics requiring
access to longer time scales,\citep{cohen_numerically_2013,cohen_taming_2015,ridley_numerically_2018}
but this is not relevant here. At each point on our plots, we display
an average of the asymptotic value of I$\left(V\right)$ and $I\left(-V\right)$
from two independent runs as our current $I$, and $\Delta I\equiv\left|I\left(V\right)-I\left(-V\right)\right|$
as a rough error estimate combining finite-time errors with stochastic
errors. In the error bars below, we bound the error estimates from
below by 3\% for the current and 5\% for the noise; and also by absolute
values of $0.01\Gamma$ and $0.01\Gamma^{2}$, respectively.

\begin{figure}
\includegraphics{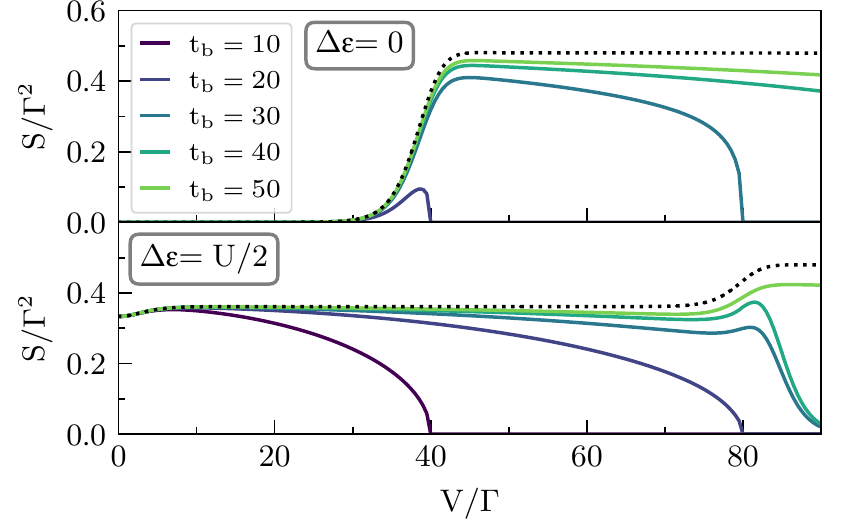}\caption{The master equation approximation to the steady state shot noise $S$
in the chain is shown as a function of bias voltage $V$, for gate
voltages at the symmetric point $\Delta\varepsilon=0$ (top) and at
$\Delta\varepsilon=\frac{U}{2}$ (bottom). Parameters are as in Fig.~\ref{fig:qme_current}.\label{fig:qme_noise}}
\end{figure}

\begin{figure}
\includegraphics{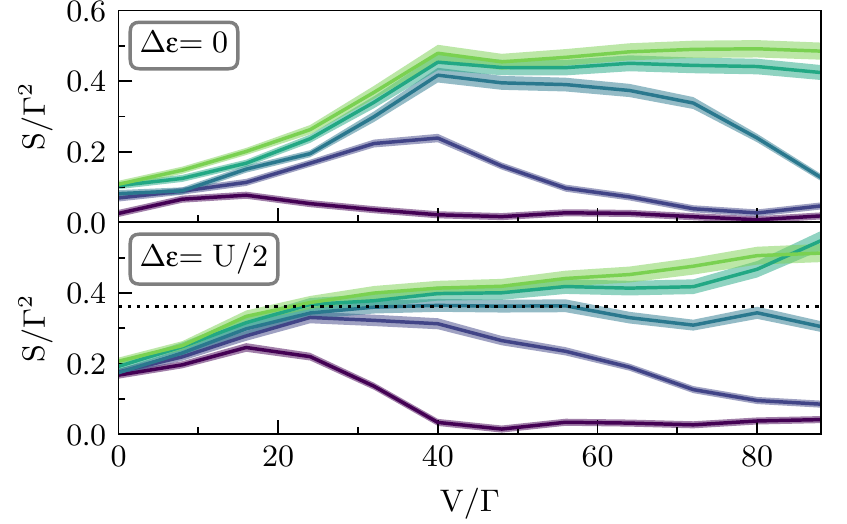}\caption{The numerically exact steady state shot noise $S$ in the chain obtained
from iQMC is shown as a function of bias voltage $V$, for gate voltages
at the symmetric point $\Delta\varepsilon=0$ (top) and at $\Delta\varepsilon=\frac{U}{2}$
(bottom). Parameters are as in Fig.~\ref{fig:qme_noise}.\label{fig:mc_noise}}
\end{figure}

A preliminary value for any data point shown below can be obtained
within minutes on a small cluster, whereas accurate high resolution
benchmarks can be obtained in hours. Here, we take a middle path between
these extremes by attempting to keep all relative errors for the current
within a range of $\sim3\%$. Since the calculations remain rather
expensive, we forgo the two-dimensional contour plot and concentrate
on the two arguably most interesting cuts through Fig.~\ref{fig:qme_contours}:
the symmetric point at $\Delta\varepsilon=0$ and the point at which
current is maximized at small voltages, $\Delta\varepsilon=\frac{U}{2}$.
This second point is also the tip of the lighter parallelogram shaped
regions that signify partial transport. Due to the same cost, Fig.~\ref{fig:mc_current}
and Fig.~\ref{fig:mc_noise} are plotted on a rougher voltage grid
than Fig.~\ref{fig:qme_current} and Fig.~\ref{fig:qme_noise}.

In Fig.~\ref{fig:qme_current} we display these two cuts across the
data, still within the master equations. The two panels (top and bottom)
show the two values of $\Delta\varepsilon$ ($\Delta\varepsilon=0$
and $\Delta\varepsilon=\frac{U}{2},$respectively). All values of
$t_{b}$ are plotted simultaneously, making it easier to identify
some of the main features. Before we point these out, we suggest that
the reader examine and compare the figure with Fig.~\ref{fig:mc_current},
which shows the corresponding iQMC data. A cursory glance reveals
both similarities and differences, meriting a discussion of the features
that takes both sets of results into account.

At large bandwidths ($t_{b}=50\Gamma$), it is clear that master equations
provide an adequate physical picture, especially at large voltages.
The main difference in going from the QME to the iQMC picture is a
broadening of features in the $I\left(V\right)$ curves. At smaller
bandwidths, where the Markovian approximation may be expected to lose
its accuracy, the differences become increasingly dramatic. The sharp
cutoff of current at the points where the bias voltage shifts the
lead bands out of resonance with each other is softened, especially
at the symmetric point $\Delta\varepsilon=0$. Whereas the master
equations predict essentially no current at any voltage at the symmetric
point for $t_{b}=10\Gamma$, the exact result shows that the current
is clearly distinguishable from zero at voltages $V\lesssim U$. For
$t_{b}=20\Gamma$, master equations predict nonzero current over only
a small voltage region near $V\approx U$, while the exact calculation
reveals currents for a wide range of voltages $0<V\lesssim2U$. While
the master equations predict a broad and robust region of negative
differential conductance at $\Delta\varepsilon=\frac{U}{2}$ at any
finite bandwidth, the exact results shift the beginning of this region
to increasingly higher voltages as $t_{b}$ increases, such that it
eventually disappears completely in the voltage range shown.

Despite all these differences, it is interesting to note that the
suppression of transport in the regime where all channels are open
at the wide band limit, as predicted by master equations, is reproduced
in the exact results\textemdash though to a lesser degree. This tells
us that while master equations may not provide us with a quantitative
picture of currents at this physical regime, they are still capable
of providing some insight into the relevant transport mechanisms.
For example, the language of discrete transport channels, though approximate,
remains useful.

The qualitative differences between the QME and iQMC data are due
to many-body correlation effects between the molecule and the leads.
Such effects are not accounted in the QME approximation. The broadening
observed here (at both large and small bandwidths) is different from
broadening due to the molecule\textendash lead coupling, which is
also not present in master equations; that effect is of order $\Gamma$,
whereas the broadening observed here is more commensurate with the
size of the interaction strength $U=40\Gamma$. Since $U$ is rather
large here (and was chosen this way because all energy scales must
be much larger than $\Gamma$ for master equations to be valid), a
low-order perturbative expansion in the many-body interaction cannot
quantitatively capture this effect.

Next, we discuss shot noise. In Fig.~\ref{fig:qme_noise} we show
the master equation approximation, while in Fig.~\ref{fig:mc_noise}
we present the corresponding iQMC data. Although the statistical errors
associated with the noise are greater than those for the current,
as discussed in Section.~\ref{subsec:Inchworm-qmc-method}, the noise
is a more sensitive probe of the available transport channels than
the current. This has several interesting implications for the correlated
dynamics that can be clearly seen in the data.

Current fluctuations can occur in conductors held at finite temperature
even without a bias voltage. The source of these fluctuations lies
in the thermal agitation of conducting electrons, and is referred
to as the thermal noise or Johnson\textendash Nyquist noise. In Fig.~\ref{fig:qme_noise}
and Fig.~\ref{fig:mc_noise}, at the leftmost point $V=0$, the Johnson\textendash Nyquist
noise is shown at the particle\textendash hole symmetric gate voltage
$\Delta\varepsilon=0$ (top) and at $\Delta\varepsilon=\frac{U}{2}$
(bottom). Within QME (Fig.~\ref{fig:qme_noise}) the thermal noise
is zero at the symmetric point, but takes on a finite value at $\Delta\varepsilon=\frac{U}{2}$.
However, the iQMC data (Fig.~\ref{fig:mc_noise}) indicates the existence
of a finite thermal noise at both gate voltages. In particular, the
thermal noise at the symmetric point is comparable in magnitude to
the finite voltage shot noise, and appears to be weakly enhanced for
larger values of $t_{b}$.

The noise also remains a more sensitive probe of many-body scattering
effects than the current at higher bias voltages $V$. In particular,
the iQMC noise (Fig.~\ref{fig:mc_noise}) shows a more significant
broadening than that in the current plots of Fig.~\ref{fig:mc_current}.
Another striking difference is that the iQMC noise rapidly exceeds
the QME plateaus, clearly indicating an opening of the higher energy
channels at lower voltage that was not visible in the current. At
the lower bandwidths and high bias, where iQMC predicts no current,
nonzero current fluctuations clearly remain. This suggests that some
transport channels are not yet fully closed, but nevertheless carry
no current on average.

\section{Conclusions\label{sec:Conclusions}}

The electronic structure of the leads plays an essential role in dictating
the conductance properties of molecular junctions, which is lost when
one only considers the wide band limit, where Markovian QME approximations
are appropriate. We presented a numerically exact study of nonequilibrium
transport in the Anderson\textendash Newns model, which describes
an interacting impurity coupled to a one dimensional chain, using
the numerically exact iQMC method. Our methodology provides not only
the expectation value of the current, but also the full counting statistics
of particle transport, and can easily be extended to any lead geometry.
We have provided a detailed discussion of how this can be done numerically
for very large leads with potentially complex structure, and discussed
the effect on the band width and shape when the leads are modified
from the one dimensional chain structure to a two dimensional square
lattice. We leave study of the interacting square lattice, and the
effect of dimensionality in general, to future work.

We found that the finite bandwidth of the chain can suppress high
energy transport channels. Whereas the QME approximation predicts
the clean closure of transport channels above some threshold voltage
set by the bandwidth, the iQMC method shows that transport continues
to exist at a larger range of voltages. Our results suggest that this
is facilitated by an interaction-induced ``smearing'' of transport
characteristics that partially opens conduction channels forbidden
within the QME approximation, and also allows transport at lower voltages.

Our results further show an enhancement of the thermal noise compared
to QME. This enhancement depends only weakly on the lead bandwidth.
At certain parts of the parameter space we explored, the mean current
is entirely suppressed by the bandwidth, but its fluctuations are
not. This suggests that some transport processes are not forbidden,
but only average to zero, and is a powerful demonstration of Landauer's
maxim that ``the noise is the signal'' in molecular electronics.\citep{landauer1998condensed}
Understanding this effect and how it might be manipulated is a promising
direction for future work.

\paragraph{Acknowledgments}

G.C. acknowledges support by the Israel Science Foundation (Grant
No. 1604/16). M.R. was supported by the Raymond and Beverly Sackler
Center for Computational Molecular and Materials Science, Tel Aviv
University. E.G. was funded by DOE Grant No. ER 46932. This collaboration
was supported by Grant No. 2016087 from the United States-Israel Binational
Science Foundation (BSF).

\bibliographystyle{apsrev4-1}
\bibliography{Library}

\end{document}